\documentclass[sigconf]{acmart}

\copyrightyear{2026}
\acmYear{2026}
\setcopyright{cc}
\setcctype{by}
\acmConference[MM '26] {Proceedings of the 34th ACM International Conference on Multimedia}{November 10--14, 2026}{Rio de Janeiro, Brazil.}
\acmBooktitle{Proceedings of the 34th ACM International Conference on Multimedia (MM '26), November 10--14, 2026, Rio de Janeiro, Brazil}
\acmISBN{979-8-4007-2213-4/2026/11}
\acmDOI{10.1145/3767308.3836237}

\AtBeginDocument{%
  }

\usepackage{tabularx}
\usepackage{booktabs}
\usepackage{array}
\usepackage{multirow}
\usepackage{pifont}
\usepackage{amsmath}

\usepackage{float}
\usepackage{bbm}
\usepackage[table]{xcolor} 
\usepackage{colortbl}
\usepackage[most]{tcolorbox} 
\definecolor{LightBlue}{RGB}{230, 242, 255}
\definecolor{myblue}{RGB}{0,70,170}
\definecolor{mainblue}{RGB}{41, 128, 185}
\definecolor{lightblue}{RGB}{235, 245, 251}
\definecolor{mainpurple}{RGB}{142, 68, 173}
\definecolor{lightpurple}{RGB}{244, 236, 247}
\definecolor{codegray}{RGB}{248, 249, 250}

\newtcolorbox{promptWithKnowledgeInfoSeek}{
  colback=white,
  colframe=mainblue,
  boxrule=1.5pt,
  arc=3mm,
  breakable,
  fonttitle=\bfseries\normalsize,
  title={\hspace{2mm}Retrieval-Augmented Prompt for InfoSeek},
  coltitle=white,
  colbacktitle=mainblue,
  left=3pt,
  right=5pt,
  top=5pt,
  bottom=5pt,
  before skip=15pt,
  after skip=15pt,
}

\newtcolorbox{promptWithKnowledgeEVQA}{
  colback=white,
  colframe=mainblue,
  boxrule=1.5pt,
  arc=3mm,
  breakable,
  fonttitle=\bfseries\normalsize,
  title={\hspace{2mm}Retrieval-Augmented Prompt for E-VQA},
  coltitle=white,
  colbacktitle=mainblue,
  left=3pt,
  right=5pt,
  top=5pt,
  bottom=5pt,
  before skip=15pt,
  after skip=15pt,
}

\newtcolorbox{promptZeroShot}{
  colback=white,
  colframe=mainpurple,
  boxrule=1.5pt,
  arc=3mm,
  breakable,
  fonttitle=\bfseries\normalsize,
  title={\hspace{2mm}Zero-Shot Prompt},
  coltitle=white,
  colbacktitle=mainpurple,
  left=3pt,
  right=5pt,
  top=5pt,
  bottom=5pt,
  before skip=15pt,
  after skip=15pt
}

\newtcolorbox{systemBox}[1][mainblue]{
  colback=#1!5,
  colframe=#1,
  boxrule=0pt,
  leftrule=3pt,
  arc=2mm,
  left=8pt,
  right=5pt,
  top=5pt,
  bottom=6pt,
  breakable,
  fontupper=\ttfamily\raggedright,
  halign=flush left,
  before skip=3pt,,
  after skip=6pt,
  title={System Prompt},
  fonttitle=\bfseries\footnotesize,
  coltitle=#1,
  colbacktitle=#1!10,
  titlerule=0pt
}

\newtcolorbox{userBox}[1][mainblue]{
  colback=codegray,
  colframe=#1,
  boxrule=0.8pt,
  arc=2mm,
  left=8pt,
  right=8pt,
  top=6pt,
  bottom=6pt,
  breakable,
  fontupper=\ttfamily,
  halign=flush left,
  before skip=8pt,
  after skip=3pt,
  title={User Prompt},
  fonttitle=\bfseries\footnotesize,
  coltitle=#1,
  colbacktitle=#1!10,
  titlerule=0pt
}

\begin{document}

\title{UniHEAR: Unified Heterogeneous-Source Attentive Retrieval for Knowledge-Based Visual Question Answering}

\author{Ganzhong Luo}
\affiliation{%
  \institution{School of Aeronautics and Astronautics, Sichuan University}
  \city{Chengdu}
  \country{China}
}
\email{luoganzhong@stu.scu.edu.cn}
% \orcid{0009-0003-5724-5551}

\author{Yang Ren}
\affiliation{%
  \institution{School of Aeronautics and Astronautics, Sichuan University}
  \city{Chengdu}
  \country{China}
}
\email{renyang@stu.scu.edu.cn}

\author{Hanyong Wang}
\affiliation{%
  \institution{School of Aeronautics and Astronautics, Sichuan University}
  \city{Chengdu}
  \country{China}
}
\email{harryw@stu.scu.edu.cn}

\author{Shuyu Zheng}
\affiliation{%
  \institution{School of Aeronautics and Astronautics, Sichuan University}
  \city{Chengdu}
  \country{China}
}
\email{slim\_zeng@stu.scu.edu.cn}

\author{Menglong Yang}
\authornote{Corresponding author.}
\affiliation{%
  \institution{School of Aeronautics and Astronautics, Sichuan University}
  \city{Chengdu}
  \country{China}
}
\email{mlyang@scu.edu.cn}

\renewcommand{\shortauthors}{Ganzhong Luo, Yang Ren, Hanyong Wang, Shuyu Zheng, and Menglong Yang}

\begin{abstract}
Knowledge-Based Visual Question Answering (KB-VQA) requires retrieving entity knowledge from external sources to answer visually grounded questions. Existing retrieval-augmented systems suffer from two critical limitations. First, relying on a single retrieval modality creates a Single-Source Retrieval Bottleneck, missing ground-truth entities that are only accessible through complementary sources. Second, dual-tower pointwise rerankers suffer from Retrieval-Source-Blind Reranking, as they overlook retrieval origins and candidate-level retrieval priors, leading to redundant modality reliance. To address these challenges, we propose UniHEAR, a unified lightweight framework for heterogeneous-source entity retrieval and reranking. UniHEAR constructs a Coarse Retrieval Descriptor for each candidate entity, and introduces Retrieval-Guided Attentive Modality Gating to condition modality attention weights on this descriptor, complemented by Entropy-Weighted Source Fusion of coarse retrieval priors. A hybrid training strategy combining contrastive learning with an auxiliary modality-preserving loss unifies entity-level and section-level retrieval within a single model. Extensive experiments on E-VQA and InfoSeek demonstrate that UniHEAR achieves state-of-the-art retrieval and VQA performance, improving Recall@1 by 6.7 and 1.2 points over the strongest baselines while maintaining a lightweight reranking architecture.
Code and model are available at https://github.com/iven-luo/UniHEAR.
\end{abstract}




\maketitle

\begin{figure}[h]
  \centering
  \vspace{-10pt}
  \includegraphics[width=\linewidth]{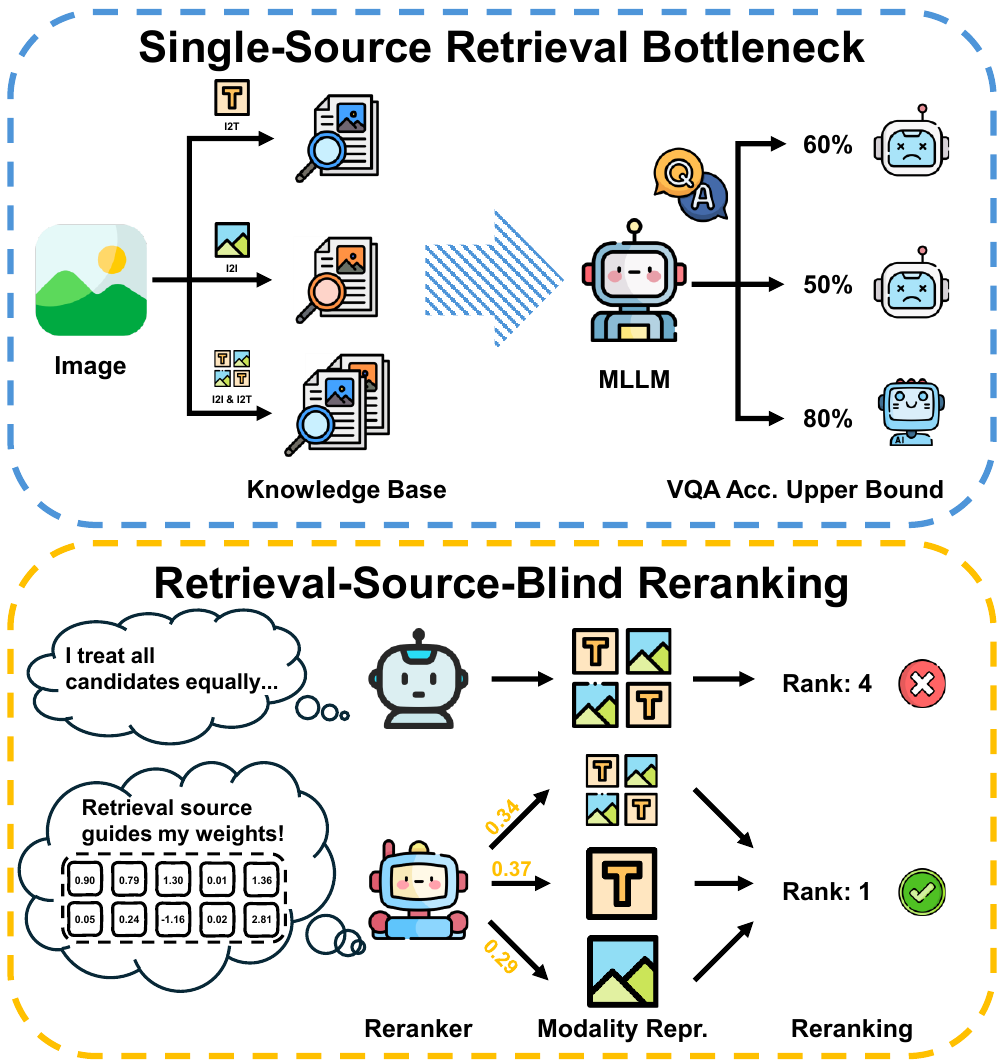}
    \caption{Limitations of existing KB-VQA systems.}
    \label{fig:Limitations}
    \vspace{-10pt}
\end{figure}

\section{Introduction}
Visual Question Answering (VQA)~\cite{Vqa} requires models to answer natural language questions about images. A more challenging variant, Knowledge-Based VQA (KB-VQA), further requires external entity knowledge beyond visual content from sources like Wikipedia. While Multimodal Large Language Models (MLLMs)~\cite{qwen3vl, LLaVA1.5, Internvl3} excel at general VQA, they often require external augmentation for knowledge-intensive queries. Consequently, Retrieval-Augmented Generation (RAG)~\cite{Wiki-llava, mr2ag, ReflectiVA, EchoSight, omgm, mKG-RAG} has become the dominant paradigm, retrieving query-relevant knowledge to contextualize MLLM generation. RAG frameworks typically adopt a three-stage pipeline where coarse 
retrieval extracts top-$k$ candidate entities via vision-language 
encoders, fine-grained reranking optimizes the candidate order, and 
answer generation derives the final response based on the top-ranked 
entity's knowledge. 

As illustrated in Figure~\ref{fig:Limitations}, current systems face two limitations. First, the reliance on a single retrieval source, either image-to-image (I2I) or image-to-text (I2T), creates a Single-Source Retrieval Bottleneck. These modalities diverge significantly: I2I prioritizes morphological similarity, while I2T emphasizes semantic consistency, yielding largely non-overlapping candidate sets under limited retrieval budgets. Consequently, ground-truth entities are frequently captured by only one source, rendering single-source retrieval inherently incomplete and suppressing the VQA accuracy upper bound. Second, existing rerankers suffer from Retrieval-Source-Blind Reranking: they assign uniform modality attention across all candidates regardless of their retrieval source, ignoring the rich signals already embedded in coarse retrieval. This leads to redundant reliance on already-exploited modalities, reducing reranking discriminability and causing misranking of ground-truth entities. Specifically, lightweight rerankers fail to exploit the rich retrieval evidence already surfaced by large-scale coarse retrievers, leaving valuable source-aware signals entirely unutilized during reranking.

To overcome these limitations, we propose \textbf{UniHEAR}, \textbf{Uni}fied \textbf{He}terogeneous-Source \textbf{A}ttentive \textbf{R}etrieval for Knowledge-Based Visual Question Answering. UniHEAR is a lightweight multimodal RAG framework that explicitly bridges coarse retrieval and fine-grained reranking via retrieval priors within a unified architecture. Specifically, we perform Heterogeneous-Source Coarse-Grained Retrieval by simultaneously matching the query image against entity images and textual summaries, constructing a unified candidate pool with complementary coverage. Each candidate is then characterized by a Coarse Retrieval Descriptor that encodes rank, similarity, and entropy statistics from both sources. Subsequently, we introduce Retrieval-Guided Attentive Modality Gating in the reranking stage, which conditions modality attention weights on the Coarse Retrieval Descriptor and aggregates modality-specific representations into a unified fused representation, enabling the lightweight reranker to exploit rich prior signals from large-scale coarse retrievers and reduce over-reliance on already-exploited modalities. Coarse retrieval priors are re-incorporated via Entropy-Weighted Source Fusion as a training-free complement to the learned reranking score. Finally, Multi-Entity Section-Augmented Generation aggregates sections across the top-$k_e$ reranked entities with rank-proportional allocation, providing comprehensive knowledge context for MLLM-based answer generation. Furthermore, we employ a training strategy that learns discriminative fused representations via contrastive learning, while an auxiliary loss preserves cross-modal matching capabilities for downstream section-level retrieval. Extensive experiments on E-VQA and InfoSeek demonstrate that our method outperforms existing state-of-the-art competitors. Our main contributions are as follows:

\begin{itemize}
    \item We propose UniHEAR, a unified lightweight framework for heterogeneous-source 
    entity retrieval and reranking in KB-VQA, which constructs a Coarse Retrieval 
    Descriptor to encode multi-source retrieval evidence as a bridge between 
    large-scale coarse retrievers and the fine-grained reranker.
    
    \item We introduce Retrieval-Guided Attentive Modality Gating, which 
    conditions modality attention on the Coarse Retrieval Descriptor to mitigate 
    redundant modality reliance and aggregates modality-specific representations 
    into a unified fused representation, complemented by Entropy-Weighted Source 
    Fusion that re-incorporates calibrated coarse retrieval priors as a 
    training-free signal.
    
    \item We design a training strategy that unifies entity- and section-level 
    retrieval within a single reranker, combining contrastive learning on fused 
    representations with an auxiliary modality-preserving loss to enable seamless 
    entity-to-section retrieval, complemented by a rank-proportional section allocation 
    strategy across multiple entities for answer generation.
    
    \item Extensive experiments on E-VQA and InfoSeek demonstrate that
    UniHEAR achieves state-of-the-art retrieval and VQA performance,
    improving Recall@1 by 6.7 and 1.2 points over the strongest baselines while maintaining a lightweight reranking architecture.
\end{itemize}

\begin{figure*}[!h]
  \centering
  \includegraphics[width=\textwidth]{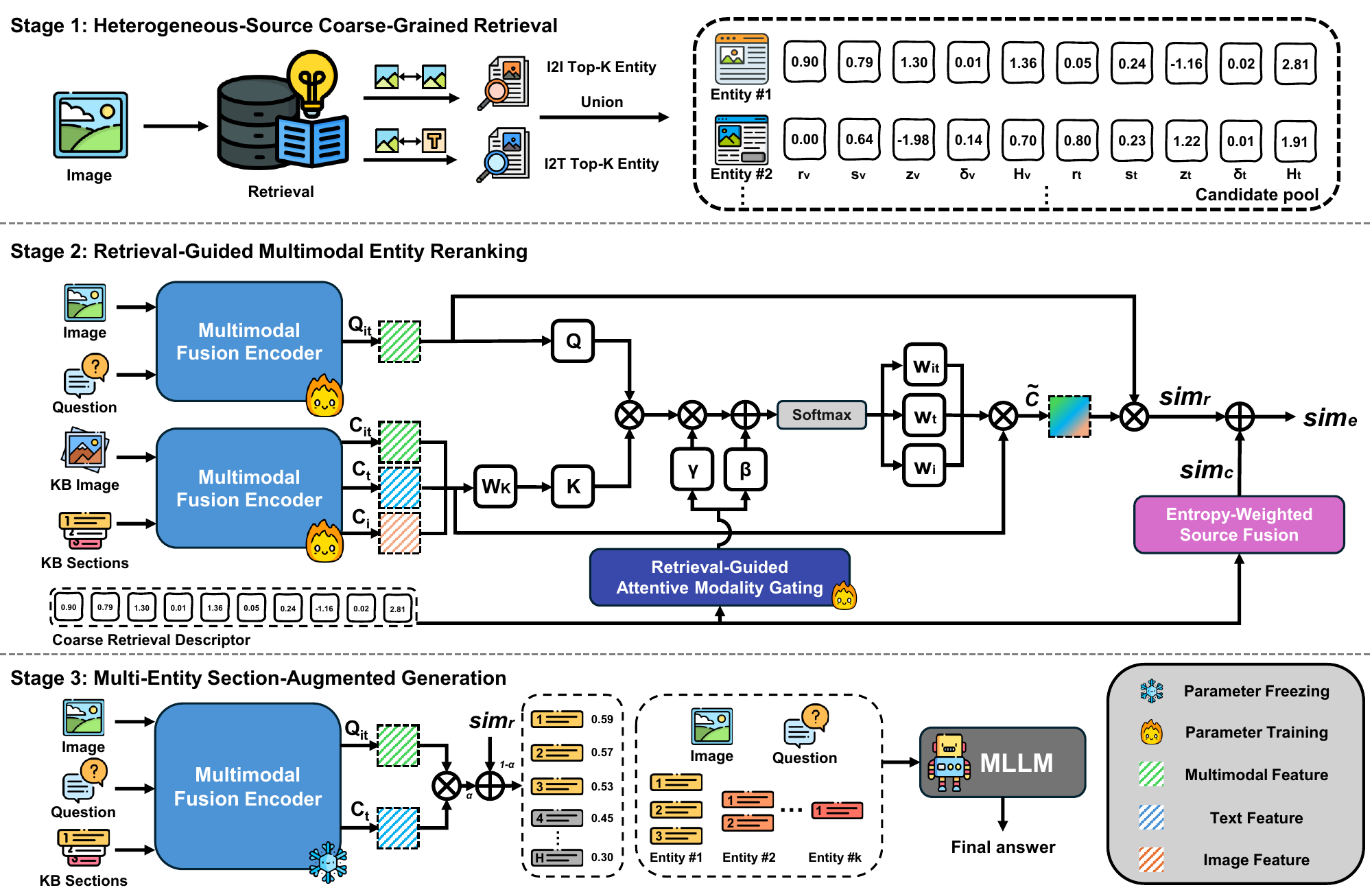}
  \caption{Overview of the UniHEAR framework.}
  \label{fig:pipeline}
\end{figure*}

\section{Related Work}
\subsection{Knowledge-Based VQA}
KB-VQA extends VQA~\cite{Vqa} by incorporating external knowledge for questions beyond visual content. 
The field has advanced significantly through new benchmarks~\cite{evqa, INFOSEEK} 
and LLM integration, with In-Context Learning approaches~\cite{PICa, Promptcap, 
MM-reasoner, Prophet} leveraging parametric knowledge in large language 
models~\cite{Gpt-4, Llama2}. However, such methods rely entirely on 
knowledge stored in model weights, lacking explicit retrieval mechanisms 
to ground predictions in verifiable evidence.
Retrieval Augmented Generation (RAG)~\cite{lin2022retrieval} has been expanded to KB-VQA to guide generation with external documents. Existing methods employ fine-grained encoding~\cite{Preflmr, MuKA}, hierarchical strategies~\cite{Wiki-llava, EchoSight}, denoising~\cite{LLM-RA, Rora-vlm}, or reflective refinement~\cite{mr2ag, ReflectiVA} to improve retrieval. Others utilize knowledge graphs~\cite{mKG-RAG} or reinforcement learning~\cite{VLM-PRF}. To address the knowledge boundaries of multimodal LLMs, ReflectiVA~\cite{ReflectiVA} introduces a self-reflective mechanism that enables the model to autonomously decide when to invoke external databases and how to integrate the retrieved content effectively. To mitigate retrieval noise and enhance reasoning in KB-VQA, ReAG~\cite{ReAG} employs a multi-level retrieval pipeline and a GRPO-inspired reinforcement learning protocol to optimize the generation of explicit reasoning traces from retrieved multimodal documents.

\subsection{Multimodal Knowledge Filtering in KB-VQA}
In KB-VQA systems, retrieval from large-scale knowledge bases is inherently imprecise: the retrieved candidates frequently contain irrelevant, redundant, or even contradictory information relative to the question at hand. Without an effective filtering or reranking stage, such noise is directly propagated to the answer generation module, where it can mislead model reasoning and substantially degrade final answer accuracy. Critically, the filtering challenge in KB-VQA is not merely textual, as it demands joint reasoning over visual content, question semantics, and heterogeneous knowledge sources, making naive similarity-based selection insufficient. A well-designed filtering mechanism therefore serves as an essential bridge between coarse retrieval and precise answer generation, determining not only which knowledge is retrieved, but which knowledge is truly useful. EchoSight~\cite{EchoSight} proposes a multimodal RAG framework that enhances knowledge-based visual question answering by employing a dual-stage retrieval and reranking mechanism to effectively integrate fine-grained encyclopedic knowledge. OMGM~\cite{omgm} specifically addresses granularity mismatch between coarse entity retrieval and fine-grained section matching. 

However, these approaches predominantly rely on a single retrieval modality during filtering or reranking, failing to exploit the multi-source signals already surfaced by large-scale coarse retrievers. As a result, the reranking stage remains constrained by the redundant and incomplete information inherited from coarse retrieval, limiting its capacity to identify the most informative candidates. In contrast, our framework explicitly leverages the complementary signals from heterogeneous retrieval sources to guide a lightweight reranker, harnessing the synergistic potential of multi-source coarse retrieval to achieve efficient and precise candidate selection.

\section{Methodology}
In this section, we present UniHEAR, a unified lightweight framework 
for heterogeneous-source entity retrieval and reranking in KB-VQA. 
As illustrated in Figure~\ref{fig:pipeline}, UniHEAR operates in
three stages. First, heterogeneous-source coarse retrieval constructs
a unified candidate pool from both visual and textual sources and
characterizes each candidate with a coarse retrieval descriptor.
Second, retrieval-guided multimodal reranking conditions modality
attention on this descriptor and re-incorporates coarse retrieval
priors through entropy-weighted source fusion. Third, multi-entity
section-augmented generation aggregates rank-proportional sections
across the top reranked entities for answer generation.

\subsection{Heterogeneous-Source Coarse-Grained Retrieval}

Coarse-grained entity retrieval aims to identify top-$k$ candidates from a multimodal knowledge base. However, retrieval modalities exhibit distinct characteristics. Visual-visual matching prioritizes morphological similarity such as color and shape, which often causes visually similar but semantically distinct entities to be conflated. Visual-text matching instead emphasizes semantic consistency, which tends to retrieve entities that share category-level semantics with the query but correspond to an incorrect specific instance. Consequently, our heterogeneous-source retrieval strategy is designed to leverage the complementary strengths of these distinct retrieval behaviors.

Beyond pooling candidates from both sources, we characterize each candidate with a Coarse Retrieval Descriptor $\mathcal{D}(e)$ that injects coarse retrieval evidence into reranking as prior guidance and, more importantly, compensates for the limited global context of dual-tower pointwise rerankers by encoding each candidate's relative standing within the retrieval distribution.
Specifically, given a query image $\mathcal{I}_q$, candidate entities are retrieved by matching the query against two complementary knowledge-base sources, including entity images $\mathcal{I}_e$ and textual summaries $\mathcal{T}_e^s$ of entity articles. We denote the source-specific knowledge-base content as $\mathcal{K}_s$ and compute coarse-grained similarities between the query and each source, formalized as,

\begin{equation}
\label{eq:sim_coarse_retrieval}
\mathrm{sim}_c^s(e) =
\frac{E_v(\mathcal{I}_q) \cdot E_s(\mathcal{K}_s)}
{\lVert E_v(\mathcal{I}_q) \rVert \lVert E_s(\mathcal{K}_s) \rVert},
\quad s \in \{v,t\}
\end{equation}
where $\mathcal{K}_s$ denotes the raw knowledge-base content for source $s$, with $\mathcal{K}_v = \mathcal{I}_e$ and $\mathcal{K}_t = \mathcal{T}_e^s$, and $E_v(\cdot)$, $E_t(\cdot)$ denote the CLIP visual and textual encoders, respectively.

We employ Faiss~\cite{Faiss} to retrieve the top-$k$ nearest neighbors via inner-product similarity $\mathrm{sim}_c^s(e)$ over pre-indexed embeddings, yielding two source-specific candidate sets $\mathcal{C}_v^{k}$ and $\mathcal{C}_t^{k}$. The final candidate pool $\mathcal{C} = \mathcal{C}_v^k \cup \mathcal{C}_t^k$ is the union of both sources, including entities captured exclusively by a single modality. For each source $s$, we compute a pool-level entropy $H_s$ to measure 
the concentration of the top-$k$ retrieval distribution, reflecting 
source reliability for downstream reranking:
\begin{equation}
H_s = - \sum_{i=1}^{k} p_s^i \log p_s^i, \quad p_s^i = \frac{\exp(\mathrm{sim}_c^s(e_i)/\tau_h)}{\sum_{j=1}^{k} \exp(\mathrm{sim}_c^s(e_j)/\tau_h)}
\end{equation}
where the temperature $\tau_h$ amplifies the differences among similarity scores, making the entropy more discriminative for downstream reranking.

For each entity $e \in \mathcal{C}$, we define the normalized rank as 
$r_s(e) = 1 - (\mathrm{rank}_s(e)-1)/k$, where $\mathrm{rank}_s(e)$ 
denotes the rank position of entity $e$ in source $s$. Entities absent 
from source $s$ are assigned rank $k+1$, yielding $r_s(e)=0$, and their 
similarity is estimated as $\min_{e \in \mathcal{C}_s^k}\,\mathrm{sim}_c^s(e) - \sigma_s$, 
where $\sigma_s$ is the standard deviation of similarities in $\mathcal{C}_s^k$. The source-specific descriptor is then
\begin{equation}
\mathbf{d}_s(e) = \big[ r_s(e),\; \mathrm{sim}_c^s(e),\; z_s(e),\; \delta_s(e),\; H_s \big]
\end{equation}
where $z_s(e)$ is the z-normalized similarity, and $\delta_s(e)$ measures the similarity gap to the top-ranked entity. The final Coarse Retrieval Descriptor for entity $e$ is obtained by concatenating both source descriptors:
\begin{equation}
\mathcal{D}(e)= \big [\mathbf{d}_v(e),\;\mathbf{d}_t(e)\big]\in \mathbb{R}^{10}
\end{equation}

\subsection{Retrieval-Guided Multimodal Entity Reranking}
Given the candidate set $\mathcal{C}$, we perform fine-grained reranking through two complementary modules. Retrieval-Guided Attentive Modality Gating (RAMG) conditions modality attention weights on the retrieval source of each candidate, mitigating redundant reliance on modalities already exploited during coarse retrieval. Entropy-Weighted Source Fusion (EWSF) then re-incorporates coarse retrieval priors, complementing the learned reranking score with rank- and entropy-weighted source evidence.

We first employ VISTA~\cite{VISTA} as the multimodal encoder. Given an image-text pair $(\mathcal{I}, \mathcal{T})$, the encoder produces three modality-specific representations:
\begin{equation}
    f_m = \text{BERT}(\phi_m(\mathcal{I},\mathcal{T})), \quad m \in \{it, t, i \}
\end{equation}
where $\phi_i$ extracts visual tokens via ViT~\cite{VIT}, $\phi_t$ processes textual tokens, and $\phi_{it}$ concatenates both. These token sequences are then encoded by BERT~\cite{Bert}.

For the query, we encode the image-question pair $(\mathcal{I}_q, \mathcal{T}_q)$ to obtain multimodal query representation $\mathbf{Q}_{it}$. For each candidate entity in $\mathcal{C}$ with knowledge base image $\mathcal{I}_e$ and article $\mathcal{T}_e$ comprising $H$ sections $\mathcal{T}_e^{h}$ ($h \in [1, H]$), following VISTA, we encode $\mathcal{I}_e$ once to obtain a shared visual representation $\mathbf{C}_i$, and pair $\mathcal{I}_e$ with each section $\mathcal{T}_e^h$ to obtain $H$ section-level multimodal and textual representations $\mathbf{C}_{it}^h$ and $\mathbf{C}_t^h$.

\textbf{Retrieval-Guided Attentive Modality Gating. }
Conventional rerankers directly aggregate multimodal features via IT2IT matching without explicit modality weighting, and are unaware of which modalities have already been exploited during coarse retrieval. 
This can lead to redundant modality usage: a candidate retrieved via visual search $\mathcal{C}_v^k$ has already been confirmed in the visual space, making further reliance on visual features less discriminative.
To better exploit the rich prior signals provided by large-scale retrieval models and guide a lightweight reranker to perform more efficient reranking, we propose Retrieval-Guided Attentive Modality Gating, which adjusts modality attention weights by conditioning the attention logits on the Coarse Retrieval Descriptor $\mathcal{D}(e)$.

Specifically, for each section $h$ of candidate entity $e$, we define the query $Q \in \mathbb{R}^{1 \times d}$ as the multimodal query representation $\mathbf{Q}_{it}$, and the key matrix $K \in \mathbb{R}^{3 \times d}$ as the projected stack of three modality representations:
\begin{equation}
    K = \mathbf{W}_K\, 
    [\mathbf{C}_{it}^h;\, \mathbf{C}_t^h;\, \mathbf{C}_i]
    \in \mathbb{R}^{3 \times d}
\end{equation}
where $\mathbf{W}_K \in \mathbb{R}^{d \times d}$ projects the modality representations into a discriminative subspace, enabling the model to learn which aspects of each modality are most relevant for query matching, beyond the fixed similarity implied by the shared encoder space. The content-level affinity between the query and each modality is then:
\begin{equation}
\label{eq:attention}
    [w_{it}^h,\, w_t^h,\, w_i^h] = 
    \text{Softmax}\!\left(
    \boldsymbol{\gamma}(\mathcal{D}) \odot 
    \frac{QK^T}{\sqrt{d}} + 
    \boldsymbol{\beta}(\mathcal{D})
    \right)
\end{equation}
where $\boldsymbol{\gamma}(\mathcal{D}), \boldsymbol{\beta}(\mathcal{D}) \in \mathbb{R}^3$ are source-conditioned affine modulation parameters derived from the Coarse Retrieval Descriptor:
\begin{equation}
    \boldsymbol{\gamma}(\mathcal{D}) = 
    \mathbf{W}_\gamma \mathcal{D}(e) + \mathbf{b}_\gamma, 
    \qquad
    \boldsymbol{\beta}(\mathcal{D}) = 
    \mathbf{W}_\beta \mathcal{D}(e) + \mathbf{b}_\beta
\end{equation}
$\boldsymbol{\gamma}$ and $\boldsymbol{\beta}$ act as per-modality scale and shift over the attention logits, injecting retrieval source awareness into the weight computation. $\mathbf{W}_\gamma$ and $\mathbf{W}_\beta$ are initialized such that $\boldsymbol{\gamma} = \mathbf{1}$ and $\boldsymbol{\beta} = \mathbf{0}$, ensuring RAMG degrades to standard content-only attention at initialization and learns the retrieval-guided modulation progressively.

The section-level fused representation is:
\begin{equation}
    \tilde{\mathbf{C}}^h = 
    w_{it}^h \mathbf{C}_{it}^h + 
    w_t^h \mathbf{C}_t^h + 
    w_i^h \mathbf{C}_i
\end{equation}
and the reranking score for entity $e$ is obtained by 
taking the maximum similarity across all sections:
\begin{equation}
    \mathrm{sim}_r(e) = 
    \max_{h \in [1,H]}\; 
    \mathrm{sim}_r^h(e) = 
    \max_{h \in [1,H]}\; 
    \mathbf{Q}_{it}^T \tilde{\mathbf{C}}^h
\end{equation}

\textbf{Entropy-Weighted Source Fusion.}
To further leverage the prior signals embedded in the coarse retrieval stage, we complement the reranking score with a unified coarse retrieval score $\mathrm{sim}_c^s(e)$. Each source $s \in \{v, t\}$ is weighted by:
\begin{equation}
    w_s(e) = \frac{r_s(e)}{H_s + \epsilon}
\end{equation}
where $r_s(e)$ is the normalized rank and $H_s$ is the pool-level entropy reflecting source reliability. $w_s(e)=0$ for absent sources, naturally unifying entities from one or both sources without branching. Since the two sources operate at different similarity scales, we calibrate textual similarities via a pool-mean ratio $R = \mu_v / \mu_t$ and define the unified coarse score as:
\begin{equation}
    \mathrm{sim}_c(e) = \frac{
        w_v(e)\cdot \mathrm{sim}_v^c(e) +
        w_t(e)\cdot R \cdot \mathrm{sim}_t^c(e)
    }{
        w_v(e) + w_t(e)
    }
\end{equation}

The final entity ranking score is:
\begin{equation}
    \mathrm{sim}(e) = \mathrm{sim}_r(e) + \mathrm{sim}_c(e)
    \label{eq:Score_Fusion}
\end{equation}

Entities are ranked by $\mathrm{sim}(e)$ in descending order, and the top-$k_e$ entities are retained as the final candidate set $\mathcal{E} = \{e^{(1)}, e^{(2)}, \ldots, e^{(k_e)}\}$ for downstream answer generation.

\subsection{Reranker Training Objective}
To train our reranker, we construct training pairs from the top-$k$ candidates of both $\mathcal{C}_v^k$ and $\mathcal{C}_t^k$. Each positive pair consists of the query with ground-truth entity images and their correct sections, while $N$ hard negatives are drawn from incorrect entities across both sources. All candidates are processed through RAMG to obtain fused section representations $\tilde{\mathbf{C}}^h$. We optimize two complementary objectives, where $\lambda$ balances their contributions. The primary contrastive loss trains the fused representation:
\begin{equation}
    \mathcal{L}_{con} = -\log \frac{\exp(\mathbf{Q}_{it}^T \tilde{\mathbf{C}}^{+}/\tau)}{\exp(\mathbf{Q}_{it}^T \tilde{\mathbf{C}}^{+}/\tau) + \sum_{n=1}^{N} \exp(\mathbf{Q}_{it}^T \tilde{\mathbf{C}}^n/\tau)}
\end{equation}
where $\tilde{\mathbf{C}}^{+}$ is the fused feature of the positive pair, $\tau$ is temperature. To prevent modality collapse and preserve IT2T matching capabilities required for downstream section-level reranking, we introduce an auxiliary loss that maintains alignment across multimodal, visual, and textual feature pairs:
\begin{equation}
    \mathcal{L}_{aux} = -\sum_{m \in \{it, t, i\} }
    \omega_m
    \log \frac{\exp(\mathbf{Q}_{it}^T \mathbf{C}_m^{+}/\tau)}{\exp(\mathbf{Q}_{it}^T \mathbf{C}^{+}_m/\tau) +\sum_{n=1}^{N} \exp(\mathbf{Q}_{it}^T \mathbf{C}_m^n/\tau)}
\end{equation}
where $\omega_m$ is a binary mask that filters out unreliable positive supervision: since the dataset provides no ground-truth image labels and knowledge base image URLs may be inaccessible, the positive visual signal is not always trustworthy. $\omega_m$ retains modality $m$ only when the positive score is sufficiently competitive against hard negatives, suppressing gradient updates from degraded or mismatched supervision:
\begin{equation}
    \omega_m = \mathbbm{1}\!\left[\mathbf{Q}_{it}^T \mathbf{C}_m^{+} > \displaystyle\max_{n}\, (\mathbf{Q}_{it}^T \mathbf{C}_m^n) - \theta\right]
\end{equation}

The overall training objective is:
\begin{equation}
    \mathcal{L} = \mathcal{L}_{con} + \lambda \mathcal{L}_{aux}
    \label{eq:loss}
\end{equation}

\subsection{Multi-Entity Section-Augmented Generation}
Given the candidate entity set $\mathcal{E} = \{e^{(i)}\}_{i=1}^{k_e}$, we rerank the most relevant sections across all entities for answer generation. For each section $\mathcal{T}_e^{h}$ of entity $e$  $\in \mathcal{E}$, we combine section-level image-text-to-text (IT2T) similarity with entity-level retrieval scores to jointly leverage fine-grained and coarse-grained signals:
\begin{equation}
    \mathrm{sim}_s^h(e) = \alpha\, \mathbf{Q}_{it}^T \mathbf{C}_t^{h}(e) + (1-\alpha)\, \mathrm{sim}_r^h(e)
\end{equation}
where $\mathbf{C}_t^{h}(e)$ is the textual feature of section $\mathcal{T}_e^{h}$ from entity $e$, and $\alpha$ balances fine-grained section specificity against entity-level confidence. 
To concentrate the context budget proportional to entity-level retrieval confidence, we allocate sections according to entity rank: the $i$-th entity $e^{(i)} \in \mathcal{E}$ is assigned $k_s - i + 1$ sections selected by $\mathrm{sim}_s^h(e)$, giving $\mathcal{S}^{(i)} = \{\mathcal{T}_{e^{(i)}}^{h}\}_{h=1}^{k_s - i + 1}$, for $i = 1, \ldots, k_e$.

The final answer is generated by an MLLM conditioned on the input image, question, and all retrieved sections:
\begin{equation}
    \mathrm{Final\ Answer} = \mathrm{MLLM}\!\left(
        \mathcal{I}_q,\; \mathcal{T}_q,\;
        \bigl\{\mathcal{S}^{(i)}\bigr\}_{i=1}^{k_e}
    \right)
\end{equation}

\section{Experiments}
\subsection{Datasets and Metrics}
\textbf{Datasets.} We conduct experiments on two widely-adopted KB-VQA datasets: E-VQA~\cite{evqa} and InfoSeek~\cite{INFOSEEK}. E-VQA comprises 1M triplets $(\mathcal{I}_q, \mathcal{T}_q, y)$, generated from 221K unique QA pairs associated with 16.7K entities, where each associated with up to five distinct entity images. Following EchoSight~\cite{EchoSight}, we evaluate on single-hop questions, which are answerable from a single Wikipedia page, yielding 4.7k evaluation samples and a knowledge base of 2M Wikipedia articles. InfoSeek contains 1.3M triplets distributed across roughly 11k Wikipedia entities, partitioned into 934k training, 73k validation, and 348k test samples. Since ground-truth answers for the test split are not publicly available, we evaluate on the validation set, which contains both unseen entities (Unseen-E) and novel questions (Unseen-Q). We adopt the standard 100k knowledge base subset from the original 6M pages, consistent with recent work~\cite{omgm}.

\textbf{Metrics.} For retrieval accuracy evaluation, we employ Recall@$k$ to measure whether the GT appears within the top-$k$ candidates, enforcing strict URL matching, i.e., the entity’s Wikipedia URL must exactly match the ground truth. For answer quality evaluation, we follow dataset-specific protocols, using BEM~\cite{Bertscore} for the E-VQA dataset and both VQA accuracy~\cite{Vqa} and Relaxed accuracy~\cite{relatex-accuracy} for InfoSeek, consistent with standard VQA evaluation practices.

\begin{table*}[t]
  \centering
  % \footnotesize
  \small
  \setlength{\tabcolsep}{8pt} 
    \begin{tabular}{lcccccccccc}  
    \toprule
    \multirow{2}{*}{\textbf{Method}} & & \multicolumn{4}{c}{\textbf{E-VQA}} & &  \multicolumn{4}{c}{\textbf{InfoSeek}} \\
    \cmidrule(){3-6} \cmidrule(){8-11} & & R@1 & R@5 & R@10 & R@20 & & R@1 & R@5 & R@10 & R@20 \\
    \midrule
    Wiki-LLaVA~\cite{Wiki-llava}  & & 3.3 & -   & 9.9  & 13.2 & & 36.9 & -    & 66.1 & 71.9 \\
    mR$^2$AG~\cite{mr2ag}         & & -   & -   & -    & -    & & 38.0 & -    & 65.0 & 71.0 \\
    LLM-RA~\cite{LLM-RA}          & & -   & -   & -    & -    & & 47.3 & 53.8 & -    & -    \\
    VLM-PRF~\cite{VLM-PRF}        & & -   & -   & -    & -    & & 54.9 & -    & -    & -    \\
    ReflectiVA~\cite{ReflectiVA}  & & 15.6 & 36.1& -   & 49.8 & & 56.1 & 77.6 & -    & \underline{86.4} \\

    \midrule
    \rowcolor{gray!15}  \multicolumn{11}{l}{\textit{Reranking on Image-to-Text Coarse Retrieval}} \\
    w/o Reranking                               & & 19.1             & 41.2               & 49.8             & 58.7             &       & 52.6             & 73.9             & 80.0              & 84.8 \\
    EchoSight$^{\dagger}$~\cite{EchoSight}       & & 34.4             & 53.2               & 57.6             & 58.7             &       & 49.4             & 75.8             & 82.2              & 84.8 \\
    OMGM~\cite{omgm}                             & & \underline{42.8} & \underline{55.7}   & \underline{58.1} & 58.7             &       & \underline{64.0} & \underline{80.8} & \underline{83.6}  & 84.8 \\
    \rowcolor{LightBlue} \multicolumn{2}{l}
    {\textbf{UniHEAR (Ours)}}                     & \textbf{43.8}    & \textbf{56.5}      & \textbf{58.2}    & \textbf{58.7}    &       & \textbf{65.2}    & \textbf{81.7}    & \textbf{84.1}     & \textbf{84.8} \\

    \midrule
    \rowcolor{gray!15}  \multicolumn{11}{l}{\textit{Reranking on Image-to-Image Coarse Retrieval}}        \\
    w/o Reranking                               & & 13.3            & 31.3            & \underline{41.0} & 48.8          &       & 45.6               & 67.1             & 73.0            & 77.9 \\
    EchoSight~\cite{EchoSight}                   & & \underline{36.5}& \underline{47.9}& 48.8             & 48.8          &       & 53.2               & \underline{74.0} & \underline{77.4}& 77.9 \\
    OMGM$^{\dagger}$ ~\cite{omgm}                & & 34.0            & 47.0            & 48.8             & 48.8          &       & \underline{57.5}   & 73.3             & 76.8            & 77.9 \\
    \rowcolor{LightBlue} \multicolumn{2}{l}
    {\textbf{UniHEAR (Ours)}}                     & \textbf{40.7}   & \textbf{48.6}   & \textbf{48.8}    & \textbf{48.8} &       & \textbf{63.5}      & \textbf{76.7}    & \textbf{77.9}   & \textbf{77.9} \\

    \midrule
    \rowcolor{gray!15}  \multicolumn{11}{l}{\textit{Reranking on Heterogeneous-Source Coarse Retrieval}}         \\
    \rowcolor{LightBlue} \multicolumn{2}{l}{}    & \textbf{49.5} & \textbf{65.9} & \textbf{68.8} & \textbf{69.9} &       & \textbf{65.2} & \textbf{83.5} & \textbf{86.9} & \textbf{88.7} \\

     \rowcolor{LightBlue} \multicolumn{2}{l}{\multirow{-2}{*}{\textbf{UniHEAR (Ours)}}}   
     & \textbf{\color{myblue}\small $\Delta$+6.7} & \textbf{\color{myblue}\small $\Delta$+10.2} & \textbf{\color{myblue}\small $\Delta$+10.7}& \textbf{\color{myblue}\small $\Delta$+11.2} & 
     & \textbf{\color{myblue}\small $\Delta$+1.2} & \textbf{\color{myblue}\small $\Delta$+2.7}  & \textbf{\color{myblue}\small $\Delta$+3.3} & \textbf{\color{myblue}\small $\Delta$+2.3}\\
    
    \bottomrule
  \end{tabular}
    \caption{Retrieval performance on the E-VQA test split and the InfoSeek validation split. Our method is shown in light blue. Best results are in bold and second-best are underlined within each setting. $^{\dagger}$ indicates reproduced results and $\Delta$ rows report the absolute improvements of UniHEAR over the best baseline.}
  \label{tab:retrieval_performance}
  % \vspace{-10pt}
\end{table*}

\begin{table*}[!t]
  \centering
  % \footnotesize
  \setlength{\tabcolsep}{7pt} 
  \small
  \begin{tabular}{lcccccccc}
    \toprule
    \multirow{2}{*}{\textbf{Model}} & \multirow{2}{*}{\textbf{Generator}} & \multirow{2}{*}{\textbf{Ret. Mode}} & \multirow{2}{*}{\textbf{Gen. FT}}  & \multirow{2}{*}{\textbf{E-VQA}} & \multicolumn{3}{c}{\textbf{InfoSeek}} \\
    \cmidrule(lr){6-8}  &   &  &   & &  \textbf{Unseen-Q} & \textbf{Unseen-E} & \textbf{Overall} \\

    \midrule 
    LLaMA-3.1-8B~\cite{llama}                                    & -                      & - & -          & 16.5             & 2.1               & 0.0                   & 0.0    \\
    Qwen2.5-VL-7B~\cite{Qwen25VL}                                & -                      & - & -          & 19.0             & 18.7              & 18.7                  & 18.7   \\
    Qwen3-VL-8B~\cite{qwen3vl}                                   & -                      & - & -          & 21.1             & 20.0              & 18.0                  & 18.9   \\
                                    
    \midrule                                        
    VLM-PRF~\cite{VLM-PRF}                                       & Qwen2.5-VL-7B          &V+T& \ding{51}  & 37.1             & 43.3              & 42.7                  & 42.8 \\
    Wiki-R1~\cite{wiki-r1}                                       & Qwen2.5-VL-7B          &V+T& \ding{51}  & 41.0             & 47.8              & 42.3                  & 44.1 \\
    ReAG~\cite{ReAG}                                             & Qwen2.5-VL-7B          &V+T& \ding{51}  & 44.9             & 48.3              & 46.2                  & 47.2 \\
          
    \rowcolor{LightBlue} \textbf{UniHEAR (Ours)}                & Qwen2.5-VL-7B          &V+T& \ding{55}  & \textbf{53.2}    & 38.0              & 37.4                  & 37.7 \\
% 38.2 37.6 37.9 
    \midrule  
    ReflectiVA~\cite{ReflectiVA}                                 & LLaMA-3.1-8B           &V+T& \ding{51}  & 35.5              & 40.4              & 39.8                  & 40.1 \\
    mKG-RAG~\cite{mKG-RAG}                                       & LLaMA-3.1-8B           &V+T& \ding{51}  & 38.4              & 41.4              & 39.6                  & 40.5 \\
    VLM-PRF~\cite{VLM-PRF}                                       & LLaMA-3.1-8B           &V+T& \ding{51}  & 36.3              & 41.3              & 40.6                  & 40.8  \\
    \rowcolor{LightBlue} \textbf{UniHEAR $^{\ddagger}$ (Ours)}  & LLaMA-3.1-8B           &V+T& \ding{51}  & \textbf{50.4}     & \textbf{44.6}     & \textbf{43.4}         & \textbf{44.1} \\
% 45.0 43.5 44.2
    \midrule    
    EchoSight$^{\dagger}$~\cite{EchoSight}                       & Qwen3-VL-8B            & V & \ding{55}  & 41.1             & 30.9              & 30.4                  & 30.6 \\
    OMGM$^{\dagger}$~\cite{omgm}                                 & Qwen3-VL-8B            & T & \ding{55}  & 46.1             & 38.9              & 37.4                  & 38.1 \\
    \rowcolor{LightBlue} \textbf{UniHEAR (Ours)}                & Qwen3-VL-8B            &V+T& \ding{55}  & \textbf{56.1}    & \textbf{39.1}     & \textbf{37.9}         & \textbf{38.6} \\
    % 38.2 37.6 37.9
    \bottomrule
  \end{tabular}
    \caption{\label{tab:vqa_results}VQA performance on the E-VQA test split and the InfoSeek validation split. Our method is shown in light blue. Ret. Mode denotes the retrieval modality used and Gen.FT indicates whether generation fine-tuning is applied. $^{\dagger}$ denotes reproduced results, and $^{\ddagger}$ denotes results using the ReflectiVA fine-tuned generator.}
    % \vspace{-10pt}
\end{table*}

\subsection{Implementation Details}
For coarse-grained retrieval, we employ EVA-CLIP-8B~\cite{eva-clip-8b} as the visual encoder for both Image-to-Image and Image-to-Text retrieval. Our reranker is built upon the VISTA~\cite{VISTA} architecture and trained on 200K samples, including 160K samples from E-VQA and 40K samples from InfoSeek. Detailed descriptions of the model architecture, training strategy, and hyperparameter configurations are provided in Appendix~\ref{sec:Additional_Implementation_Details}.

\subsection{Main Results}
The results of our method compared with other approaches are presented in Table~\ref{tab:retrieval_performance} and Table~\ref{tab:vqa_results} with additional efficiency comparison in Table~\ref{tab:efficiency}.

\textbf{Retrieval Performance.}
We compare our method against state-of-the-art baselines,
including MLLM-based inference-time retrieval methods
(Wiki-LLaVA~\cite{Wiki-llava}, mR$^2$AG~\cite{mr2ag}, LLM-RA~\cite{LLM-RA},
VLM-PRF~\cite{VLM-PRF}, ReflectiVA~\cite{ReflectiVA}) and dedicated
reranker-based methods (EchoSight~\cite{EchoSight}, OMGM~\cite{omgm}). As
illustrated in Table~\ref{tab:retrieval_performance}, UniHEAR
substantially outperforms both categories. Against the strongest
MLLM-based baseline, ReflectiVA, UniHEAR achieves absolute Recall@$1$
gains of +33.9\% on E-VQA and +9.1\% on InfoSeek; against the strongest dedicated reranker, OMGM, the gains are +6.7\% and +1.2\%, respectively. UniHEAR further
outperforms all single-source baselines on their native sources with a
significantly more lightweight architecture. This advantage is
particularly pronounced on E-VQA's 2M-entity knowledge base, where under
limited candidate budgets, ground-truth entities are frequently captured
by only one modality, making heterogeneous-source retrieval critical for
maximizing recall. These consistent gains across MLLM-based
inference-time retrieval methods, dedicated reranker-based methods, and both
single- and heterogeneous-source settings collectively demonstrate the
overall superiority of our retrieval and reranking design.

\textbf{VQA Performance.}
As shown in Table~\ref{tab:vqa_results}, we compare UniHEAR against 
zero-shot vision-language models including LLaMA-3.1-8B~\cite{llama}, Qwen2.5-VL-7B~\cite{Qwen25VL}, 
and Qwen3-VL-8B~\cite{qwen3vl}, as well as a diverse set of retrieval-augmented baselines. 
Among these, Wiki-R1~\cite{wiki-r1}, ReAG~\cite{ReAG}, ReflectiVA~\cite{ReflectiVA}, and mKG-RAG~\cite{mKG-RAG} leverage large 
generative models for knowledge filtering and answer generation, while 
EchoSight~\cite{EchoSight} and OMGM~\cite{omgm} adopt lightweight rerankers under a retrieval-augmented 
paradigm most directly comparable to ours.
Without any generation fine-tuning, UniHEAR with Qwen2.5-VL-7B achieves 
53.2\% on E-VQA, surpassing all fine-tuned retrieval-augmented baselines 
under the same generator, including ReAG with an improvement of 8.3\%, demonstrating that 
our heterogeneous-source retrieval and reranking pipeline alone provides 
sufficiently high-quality knowledge context for accurate answer generation.
Under the ReflectiVA fine-tuned generator setting, UniHEAR achieves 
50.4\% on E-VQA and 44.1\% on InfoSeek, outperforming all baselines 
including mKG-RAG by +12.0\% and +3.6\% respectively.
For the most direct comparison with reranker-based methods, UniHEAR 
with Qwen3-VL-8B outperforms OMGM by +10.0\% and EchoSight by +15.0\% 
on E-VQA, and surpasses EchoSight by +8.0\% on InfoSeek, directly 
validating that heterogeneous-source retrieval with retrieval-guided 
reranking provides substantially richer and more accurate knowledge 
than single-source approaches.

\begin{table}[t]
  \centering
  \small
  % \footnotesize
  \setlength{\tabcolsep}{4.5pt} 
  \begin{tabular}{lccccccc}
    \toprule
    \textbf{Model}   & \textbf{Backbone}     & \textbf{Params}        & \textbf{ERT}             & \textbf{SRT}              & \textbf{GIT}   & \textbf{Total}\\
    \midrule
    EchoSight        & BLIP-2       & 1.2B          & \multicolumn{2}{c}{1.09}          & 1.25 & 2.34\\
    OMGM             & BLIP-2+BGE-R & 1.7B          & 2.20           & 0.39            & 1.07 & 3.66\\
    \midrule
    \rowcolor{LightBlue} \textbf{UniHEAR} & VISTA        & \textbf{197M} & \multicolumn{2}{c}{\textbf{0.97}} & \textbf{1.46}  & \textbf{2.43}\\
    \bottomrule
  \end{tabular}
    \caption{Efficiency comparison. ERT, SRT, and GIT denote the average per-sample times in seconds for Entity Reranking, Section Reranking, and Generator Inference, respectively.}
\label{tab:efficiency}
\vspace{-20pt}
\end{table}

\textbf{Efficiency Analysis.}
Table~\ref{tab:efficiency} highlights the efficiency of UniHEAR. Despite unifying heterogeneous-source retrieval, encoding three modality-specific representations ($\mathbf{C}_{it}$, $\mathbf{C}_{t}$, and $\mathbf{C}_{i}$) for each candidate, and handling a candidate pool twice the size of single-source retrieval, UniHEAR maintains competitive inference efficiency with a substantially smaller parameter footprint than BLIP-2-based competitors, whose parameter sizes range from 1.2B to 1.7B. Moreover, the proposed RAMG module introduces only 0.59M additional parameters, accounting for merely 0.3\% overhead over the VISTA backbone. This demonstrates that retrieval-guided modality gating can effectively enhance heterogeneous-source retrieval with negligible parameter cost. Importantly, the trimodal encoding is not redundant: the consistent gains from introducing attentive modality gating over all three representations confirm that each modality contributes discriminative cues beyond IT2IT matching alone, validating that the additional encoding cost translates directly into retrieval gains. Furthermore, unlike OMGM, which relies on a standalone bge-reranker-v2-m3~\cite{bge-reranker} for section reranking, both EchoSight and UniHEAR obtain entity- and section-level reranking within a single unified model, making UniHEAR a more practical and parameter-efficient solution for retrieval over large-scale knowledge bases.

\subsection{Ablation Study}
We conduct ablation studies from five perspectives to validate the
design choices of UniHEAR: (1) the effect of heterogeneous-source
retrieval, (2) the contribution of each pipeline stage to VQA accuracy,
(3) the impact of entity reranking components, (4) the contribution of
each Coarse Retrieval Descriptor component, and (5) the effectiveness
of section-level retrieval and training strategy.

\begin{figure}[h]
  \centering
  \includegraphics[width=\linewidth]{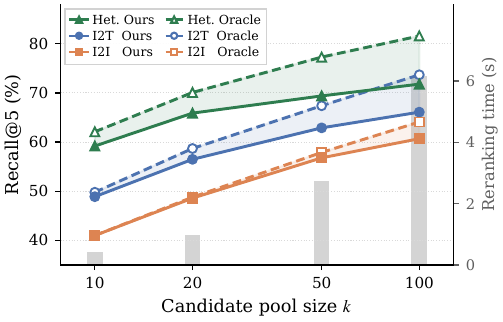}
  \caption{Analysis of heterogeneous-source retrieval under varying candidate pool sizes $k$. Oracle denotes the upper bound with perfect reranking. Gray bars indicate the average heterogeneous-source reranking time per sample. }
  \label{fig:k_ablation}
  \vspace{-5pt}
\end{figure}

\textbf{Effect of Heterogeneous-Source Retrieval.}
Figure~\ref{fig:k_ablation} examines heterogeneous-source retrieval 
under varying candidate pool sizes $k$ on E-VQA. The oracle results 
reveal the inherent ceiling of each retrieval setting: at $k{=}20$, 
the heterogeneous-source oracle achieves 70.1\%, substantially exceeding 
I2T with 58.7\% and I2I with 48.8\% alone, corroborating the single-source 
bottleneck illustrated in Figure~\ref{fig:Limitations}. UniHEAR 
consistently approaches the oracle across all settings and pool sizes, 
demonstrating effective exploitation of heterogeneous source 
complementarity through RAMG and Entropy-Weighted Source Fusion. 
While larger $k$ raises the oracle ceiling, it also increases reranking 
difficulty and inference latency with diminishing returns, as reflected in the reranking time bars. We therefore adopt $k{=}20$ as the default, 
balancing recall and efficiency.

% 四个stage的VQA准确率
\begin{table}[h]
  \centering
  \small
  % \footnotesize
  % \setlength{\tabcolsep}{4pt} % 缩小列间距
  % \renewcommand{\arraystretch}{0.9} % 缩小行距
  \begin{tabular}{ccccccc}
    \toprule
    \multirow{2}{*}{\textbf{Stage 1}}    & \multirow{2}{*}{\textbf{Stage 2}}   & \multirow{2}{*}{\textbf{Stage 3}}      & \multirow{2}{*}{\textbf{E-VQA}}     &\multicolumn{3}{c}{\textbf{InfoSeek}} \\
     \cmidrule(lr){5-7} &  & &  & \textbf{U-Q} & \textbf{U-E} & \textbf{All} \\
      \midrule      
     -         & -         & -                & 21.1               & 20.0   & 18.0   & 18.9\\
     \ding{51} & -         & -                & 28.4               & 31.9   & 31.3   & 31.6\\
     \ding{51} & \ding{51} & -                & 51.3               & 36.6   & 37.7   & 37.1 \\
     \rowcolor{LightBlue} \ding{51} & \ding{51} & \ding{51}        & \textbf{56.1}      & \textbf{39.1}   & \textbf{37.9}     & \textbf{38.6}   \\
    \bottomrule
  \end{tabular}
  \caption{Progressive ablation on VQA accuracy.}
  \label{tab:stage_ablation}
  \vspace{-10pt}
\end{table}

\textbf{Effect of Pipeline Stages.}
Table~\ref{tab:stage_ablation} presents a progressive ablation over the 
three stages of UniHEAR. The zero-shot baseline uses no external knowledge. 
Introducing I2T coarse retrieval and feeding the top-1 entity directly to 
the MLLM already yields substantial gains, confirming the necessity of 
external knowledge augmentation. Expanding the candidate 
pool to both I2I and I2T sources and applying RAMG and EWSF for heterogeneous-source 
entity reranking achieves the largest single-stage improvement, demonstrating that 
accurate entity selection across complementary sources is critical for VQA accuracy. 
Finally, multi-entity section-augmented generation further improves 
performance by providing richer and more precisely targeted knowledge 
context, with the complete pipeline achieving the best results on both 
E-VQA and InfoSeek.

% 2.实体级重排序影响
\begin{table}[h]
    \centering
    \small
  \begin{tabular}{cccccc}
    \toprule
    \multicolumn{2}{c}{\textbf{RAMG}} &  \multirow{2}{*}{\textbf{EWSF}} & \multicolumn{3}{c}{\textbf{R@1 on E-VQA}} \\ 
    \cmidrule(lr){1-2} \cmidrule(lr){4-6} \textbf{Attn. Gating} & \textbf{Ret. Guided}   & & \textbf{I2I} & \textbf{I2T} & \textbf{Het.} \\ 

    \midrule 
    -           &-           &-          &35.7       &37.5   &41.6           \\
    \ding{51}   &-           &-          &36.7       &38.7   &42.8           \\
    \ding{51}   &\ding{51}   &-          &39.6       &41.7   &46.7          \\
    \rowcolor{LightBlue} \ding{51}   &\ding{51}   &\ding{51}  &\textbf{40.7}       &\textbf{43.8}   &\textbf{49.5}           \\
    \bottomrule
  \end{tabular}
    \caption{Ablation study on entity reranking components. RAMG is decomposed into Attentive Modality Gating and Retrieval-Guided affine modulation, and EWSF denotes Entropy-Weighted Source Fusion.}
  \label{tab:reranking_entity}
  \vspace{-6pt}
\end{table}

\textbf{Effect of Entity Reranking. }
Table~\ref{tab:reranking_entity} presents a progressive ablation of the 
entity reranking components. The baseline uses IT2IT similarity for reranking 
without modality gating. Introducing trimodal attentive gating over 
$\mathbf{C}_{it}$, $\mathbf{C}_t$, and $\mathbf{C}_i$ brings consistent 
gains across all settings, confirming that multimodal representations provide 
complementary discriminative cues. Enabling retrieval-guided affine 
modulation yields substantial additional improvements, demonstrating that 
conditioning modality attention on the Coarse Retrieval Descriptor 
reduces over-reliance on already-exploited modalities. Finally, incorporating 
Entropy-Weighted Source Fusion achieves the best performance across all three 
settings, validating that coarse retrieval priors provide reliable complementary 
evidence that strengthens the learned reranking score.

\newcommand{\drop}[1]{\textcolor{red}{\footnotesize $-$#1}}
\begin{table}[h]
  \centering
  % \small
  \footnotesize
  \setlength{\tabcolsep}{6pt} % 缩小列间距
  \begin{tabular}{lccccccc@{\hspace{6pt}}c}
    \toprule
    \multirow{2}{*}{\boldmath{$r_s$}} & \multirow{2}{*}{\boldmath{$\mathrm{sim}_c^s$}} & \multirow{2}{*}{\boldmath{$z_s$}} & \multirow{2}{*}{\boldmath{$\delta_s$}} & \multirow{2}{*}{\boldmath{$H_s$}} & \multicolumn{4}{c}{\textbf{Recall@1 on E-VQA}}  \\ 
    \cmidrule(lr){6-9} & & & & & \textbf{I2I} & \textbf{I2T} & \textbf{Het.}  & \textbf{$\Delta$Het.} \\
    
      \midrule      
                          -         & -         & -         & -         &  -              & 35.7          & 37.5          & 41.6           & \drop{5.1}   \\
                          -         & \ding{51} & \ding{51} & \ding{51} &  \ding{51}      & 38.1          & 40.1          & 44.3           & \drop{2.4} \\
                          \ding{51} & -         & \ding{51} & \ding{51} &  \ding{51}      & 38.3          & 40.7          & 45.2           & \drop{1.5}   \\
                          \ding{51} & \ding{51} & -         & \ding{51} &  \ding{51}      & 37.0          & 39.1          & 43.8           & \drop{2.9}   \\
                          \ding{51} & \ding{51} & \ding{51} & -         &  \ding{51}      & 37.5          & 40.1          & 44.2           & \drop{2.5}   \\
                          \ding{51} & \ding{51} & \ding{51} & \ding{51} &  -              & 37.7          & 39.6          & 44.0           & \drop{2.7}   \\
   \rowcolor{LightBlue}   \ding{51} & \ding{51} & \ding{51} & \ding{51} &  \ding{51}      & \textbf{39.6} & \textbf{43.8} & \textbf{46.7}  & -            \\
    \bottomrule
  \end{tabular}
  \caption{Ablation of the CRD components.
$\Delta$Het.\ denotes the Recall@1 drop relative to the full CRD.}
  \label{tab:CRD_ablation}
  \vspace{-5pt}
\end{table}

\textbf{Effect of Coarse Retrieval Descriptor Component.}
Table~\ref{tab:CRD_ablation} ablates each component of the Coarse
Retrieval Descriptor (CRD) by individually removing $r_s$,
$\mathrm{sim}_c^s$, $z_s$, $\delta_s$, and $H_s$ from $\mathcal{D}(e)$.
Removing any single component degrades performance across all three
settings, confirming that each contributes independently. The largest
drops occur when $z_s$ or $H_s$ is removed, indicating that scale-invariant relative standing
and source-reliability estimation are the most critical signals, while
removing $\mathrm{sim}_c^s$ yields the smallest drop. Despite
this mild redundancy between $\mathrm{sim}_c^s$ and $z_s$, the full CRD
achieves the best overall performance across I2I, I2T, and
heterogeneous-source retrieval, and each component yields a
non-negligible gain.

% section级和融合的影响
\begin{table}[h]
  \centering
  \small
  \begin{tabular}{lccc}
    \toprule
    \textbf{Model} &  \textbf{R@1} & \textbf{R@5} & \textbf{R@10} \\

    \midrule \rowcolor{gray!15} \multicolumn{4}{l}{\textit{Image-to-Text Coarse Retrieval}}                           \\
    OMGM                                          & 32.8           &  -                 &  -    \\
    \rowcolor{LightBlue}\textbf{UniHEAR (Ours)}   & \textbf{36.2}  & \textbf{43.2}      &  \textbf{43.7}   \\

    \midrule \rowcolor{gray!15} \multicolumn{4}{l}{\textit{Image-to-Image Coarse Retrieval}}                       \\
    EchoSight$^{\dagger}$                         & 6.6            & 19.3               & 20.7    \\
    \rowcolor{LightBlue}\textbf{UniHEAR (Ours)}   & \textbf{33.1}  & \textbf{40.0}      & \textbf{40.5}   \\
  
    \midrule \rowcolor{gray!15} \multicolumn{4}{l}{\textit{Heterogeneous-Source Coarse Retrieval}}                      \\
    \textbf{UniHEAR (w/o $\mathcal{L}_{aux}$)}    & 36.6           &  42.9              &  43.4        \\
    \rowcolor{LightBlue}\textbf{UniHEAR (Ours)}   & \textbf{38.2}  &  \textbf{45.1}     &  \textbf{45.5}   \\
    \bottomrule
  \end{tabular}
  \caption{Section-level retrieval performance on E-VQA. $^{\dagger}$ indicates reproduced results.}
  \label{tab:section_reranking}
  \vspace{-5pt}
\end{table}

\textbf{Effect of Section-Level Retrieval.}
Our framework achieves unified entity-to-section retrieval within a single 
model by preserving cross-modal alignment through the auxiliary loss 
$\mathcal{L}_{aux}$, without requiring a standalone 
section reranker as in OMGM~\cite{omgm}. As shown in 
Table~\ref{tab:section_reranking}, UniHEAR consistently outperforms baselines 
under both I2I-based and I2T-based retrieval settings. Notably, ablating 
$\mathcal{L}_{aux}$ under the dual-source setting leads to consistent 
performance drops across all metrics, confirming that the auxiliary loss 
is essential for preserving the cross-modal matching capability required 
for section-level retrieval.

\section{Conclusion}
In this paper, we present UniHEAR, a unified lightweight framework that
addresses two critical limitations in KB-VQA systems: the Single-Source
Retrieval Bottleneck and Retrieval-Source-Blind Reranking. By
simultaneously querying both visual (I2I) and textual (I2T) sources,
UniHEAR constructs a unified candidate pool with complementary coverage.
A Coarse Retrieval Descriptor encodes each candidate's rank, confidence,
and distributional statistics across sources as an explicit bridge
between large-scale coarse retrievers and the fine-grained reranker,
upon which Retrieval-Guided Attentive Modality Gating conditions
modality attention to suppress already-exploited modalities.
Entropy-Weighted Source Fusion further re-incorporates coarse retrieval
priors as a training-free complement, and a hybrid training strategy
unifies entity-level and section-level retrieval within a single
lightweight model. Extensive experiments on E-VQA and InfoSeek
demonstrate that UniHEAR achieves state-of-the-art retrieval
performance, with absolute Recall@1 gains of +6.7\% and +1.2\% over the
best single-source baselines, alongside competitive VQA accuracy using a
197M-parameter reranker, 6--9$\times$ fewer parameters than competing
BLIP-2-based rerankers. We hope this work establishes heterogeneous-
source retrieval with retrieval-guided reranking as a promising paradigm
for future KB-VQA research.

\newpage

\begin{acks}
This work was supported in part by the National Natural Science Foundation of China (NSFC) under Grant No. 62271334.
\end{acks}

\bibliographystyle{ACM-Reference-Format}
\bibliography{samples/sample-base}

\clearpage
\appendix

\section*{Appendix}
\vspace{5pt}
\section{Prompt Templates}
\label{sec:prompts}
% \vspace{10pt}

We adopt three prompt formulations for answer generation. The zero-shot 
prompt instructs the model to answer solely from visual content and 
parametric knowledge, serving as a baseline. The two retrieval-augmented 
prompts are tailored to the specific evaluation protocols of E-VQA and 
InfoSeek respectively, enforcing concise answers and standardized 
numerical formatting. Both retrieval-augmented prompts explicitly 
instruct the model to leverage retrieved Wikipedia context while 
falling back to internal knowledge when the context is insufficient, 
handling three scenarios: correct entity with correct section, correct 
entity with incorrect section, and incorrect entity retrieval. This 
design prevents over-reliance on potentially irrelevant context and 
enables robust answer generation across all retrieval outcomes.

\begin{promptZeroShot}
\begin{systemBox}[mainpurple]
You are an encyclopedic visual question answering assistant. \\[2mm]

Use only internal world knowledge. 
No reasoning, no explanation. 
Output only the factual answer in no more than 5 words.
\end{systemBox}

\begin{userBox}[mainpurple]
- Question: \{question\}

Short Answer:
\end{userBox}
\end{promptZeroShot}

\begin{promptWithKnowledgeEVQA}
\begin{systemBox}[mainblue]
You are an encyclopedic visual question answering assistant.\\[2mm]

Given an image, a question, and a provided Context, directly output the answer using the Context. 
Do not mention the visual content of the image in your output.

If the context does not contain the information required to answer the question, you should answer the question using your own internal world knowledge.
\end{systemBox}

\begin{userBox}[mainblue]
- Context:  \{Wikipedia\}

- Question: \{question\}

The answer is:
\end{userBox}
\end{promptWithKnowledgeEVQA}

\begin{promptWithKnowledgeInfoSeek}
\begin{systemBox}[mainblue]
You are an encyclopedic visual question answering assistant.\\[2mm]

Answer the encyclopedic question about the given image. Don’t mention the visual content of the image in your output. Directly output the answer of the question according to the context.

If you need to answer questions about numbers or time, please output the corresponding numerical format directly. If the context does not contain the information required to answer the question, you should answer the question using internal model knowledge.

There is an example: 

- Context: 

\# Wiki Article: Dolomites 

\#\# Section Title: Dolomites The Dolomites, also known as the Dolomite Mountains, Dolomite Alps or Dolomitic Alps, are a mountain range located in northeastern Italy. The Dolomites are located in the regions of Veneto, Trentino-Alto Adige/Südtirol and Friuli Venezia Giulia, covering an area shared between the provinces of Belluno, Vicenza, Verona, Trentino, South Tyrol, Udine and Pordenone. 

- Question: Which city or region does this mountain locate in? 

Just answer the questions, no explanations needed. Short answer is: Province of Belluno 
\end{systemBox}

\begin{userBox}[mainblue]
- Context:  \{Wikipedia\}

- Question: \{question\}

Just answer the questions, no explanations needed. Short answer is:
\end{userBox}
\end{promptWithKnowledgeInfoSeek}

\section{Additional Implementation Details}
\label{sec:Additional_Implementation_Details}

\textbf{Coarse Retrieval.} For coarse-grained retrieval, we adopt 
EVA-CLIP-8B~\cite{eva-clip-8b} as the visual encoder for both I2I and I2T 
retrieval. Knowledge base embeddings are pre-indexed using Faiss~\cite{Faiss}, 
following the same construction protocol as prior works~\cite{EchoSight, omgm}.

\textbf{Reranker Architecture.} Our reranker builds upon the 
VISTA~\cite{VISTA} architecture, combining a ViT-based visual encoder 
(EVA-CLIP-02-Base~\cite{Eva-clip}) and a BERT-based text encoder 
(BGE-Base-v1.5~\cite{BGE}) for unified multimodal representation. We 
initialize from the publicly available VISTA Stage 1 checkpoint, freeze 
the visual encoder, and fine-tune the text encoder alongside the proposed 
RAMG affine modulation parameters. 
This strategy preserves pre-trained visual representations while enabling 
task-specific adaptation of textual and retrieval-guided fusion components.

\begin{table}[t]
  \centering
  % \footnotesize
  % \small
  \begin{tabular}{cccccc}
    \toprule
    \textbf{Datasets} & \textbf{Train}  & \textbf{Valid}& \textbf{Test} & \textbf{KB Size} \\
    \midrule
    E-VQA         & 160k            &11,696& 4,750 & 2M    \\
    InfoSeek      & 40k             &- & 71,335   & 100k  \\
    \bottomrule
  \end{tabular}
  \caption{Dataset statistics and knowledge base configurations.}
  \label{tab:dataset}
  \vspace{-10pt}
\end{table}
\textbf{Training.} We construct 200k training samples: 160k from 
E-VQA and 40k from InfoSeek, as detailed in Table~\ref{tab:dataset}. Since InfoSeek lacks ground-truth section 
annotations, we use jina-reranker-v3~\cite{jina} to identify the most 
relevant sections for each ground-truth entity. For each sample, we construct one positive pair and $N{=}16$ hard negatives, 
comprising 8 negatives sampled from the I2I candidates 
$\mathcal{C}_v^{k}$ and 8 from the I2T candidates 
$\mathcal{C}_t^{k}$, ensuring balanced source coverage during training.
The query consists of the input image $\mathcal{I}_q$ and question $\mathcal{T}_q$. 
Positive pairs combine the ground-truth entity's first image with its correct 
section, while hard negatives pair the first image of each negative entity 
with a randomly sampled section. We train with a learning rate of 
2e-5, global batch size of 32, using DeepSpeed ZeRO-2 
on 4$\times$ NVIDIA A6000 GPUs for approximately 6 hours per epoch. 

\textbf{Hyperparameters.} We set the coarse retrieval top-$k{=}20$ 
per source, entropy temperature $\tau_h{=}0.008$, contrastive temperature 
$\tau{=}0.02$, section scoring weight $\alpha{=}0.2$, auxiliary loss weight 
$\lambda{=}0.1$, auxiliary mask threshold $\theta{=}0.1$, and top-$k_e{=}5$ 
reranked entities each contributing top-$k_s{=}5$ sections for answer generation.
The impact of different hyperparameter settings is investigated in Appendix~\ref{sec:hyperparam_ablation}.

\section{More Ablation Experiments}
\subsection{Fairness Analysis}
\begin{table}[h]
    \centering
    % \small
    % \setlength{\tabcolsep}{5pt}
  \begin{tabular}{cccccccc}
    \toprule
    \multirow{2}{*}{\textbf{Encoder}} & \multirow{2}{*}{\textbf{Params}} & \multicolumn{3}{c}{\textbf{E-VQA}} & \multicolumn{3}{c}{\textbf{InfoSeek}} \\ 
    \cmidrule(lr){3-5} \cmidrule(lr){6-8} & & \textbf{I2I} & \textbf{I2T} & \textbf{Het.} & \textbf{I2I} & \textbf{I2T} & \textbf{Het.} \\ 

    \midrule 
    BLIP-2            &1.2B       &35.9       &40.2   &44.5     &47.9 &51.3 &47.2    \\
    VISTA             &196M       &35.7       &37.5   &41.6     &40.6 &45.5 &38.8 \\

    \bottomrule
  \end{tabular}
  \caption{Base Recall@1 of BLIP-2 and VISTA encoders fine-tuned under 
    the same IT2IT setting without reranking.}
  \label{tab:encoder_comparison}
  \vskip -10pt
\end{table}

\textbf{Multimodal Encoder Fairness Analysis.} Table~\ref{tab:encoder_comparison} compares the base retrieval recall 
of BLIP-2 and VISTA encoders under the same IT2IT fine-tuning setting, 
without any reranking. Despite VISTA achieving substantially lower base 
recall than BLIP-2 across all settings on both benchmarks, reflecting 
its significantly smaller parameter count of 196M versus 1.2B, UniHEAR 
with VISTA ultimately surpasses EchoSight and OMGM by large margins 
in the final reranking results shown in Table~\ref{tab:retrieval_performance}. 
This comparison provides two key insights. First, the performance gains 
of UniHEAR over BLIP-2-based methods cannot be attributed to encoder 
superiority; rather, they arise entirely from the heterogeneous-source 
retrieval strategy and retrieval-guided reranking. Second, the fact that 
UniHEAR substantially outperforms single-source baselines even from a 
weaker base recall demonstrates the robustness and effectiveness of 
unifying complementary retrieval sources within a lightweight framework.

\begin{table}[h]
  \centering
  % \small
  \footnotesize
  % 3.7pt
  \setlength{\tabcolsep}{2.5pt} % 缩小列间距
  \begin{tabular}{lccccccc}
    \toprule
    \textbf{Model} & \textbf{Ret.Mode}          & \textbf{ERT}$\downarrow$  & \textbf{Total}$\downarrow$    & \textbf{R@1}$\uparrow$  & \textbf{R@5}$\uparrow$  & \textbf{R@10}$\uparrow$  & \textbf{Acc.}$\uparrow$ \\

      \midrule      
     EVA-CLIP-8B                                &  V        & -                 & -                 & 13.3          & 31.3          & 41.0           & 26.6  \\
     EchoSight                                  &  V        & 1.09              & 2.34              & 36.5          & 47.9          & 48.8           & 41.1  \\
     \rowcolor{LightBlue}\textbf{UniHEAR(Ours)} &  V        & \textbf{0.47}     & \textbf{1.93}     & \textbf{40.7} & \textbf{48.6} & \textbf{48.8}  & \textbf{44.9} \\
     \midrule                           
     EVA-CLIP-8B                                &  T        & -                 & -                 & 19.1          & 41.2          & 49.8           & 28.4 \\
     OMGM                                       &  T        & 2.20              & 3.66              & 42.8          & 55.7          & 58.1           & 46.1  \\
     \rowcolor{LightBlue}\textbf{UniHEAR(Ours)} &  T        & \textbf{0.50}     & \textbf{1.96}     & \textbf{43.8} & \textbf{56.5} & \textbf{58.2}  & \textbf{46.7} \\
     \midrule                           
     EVA-CLIP-8B                                &  V+T      & -                 & -                 & 21.4          & 45.2          & 55.2           & 30.4 \\  
     \rowcolor{LightBlue}\textbf{UniHEAR(Ours)} &  V+T      & \textbf{0.97}     & \textbf{2.43}     & \textbf{49.5} & \textbf{65.9} & 68.8           & \textbf{56.1} \\
     \quad w/o RAMG\&EWSF                       &  V+T      & -                 & -                 & 41.6          & 62.3          & \textbf{69.9}  & 45.9 \\
     \quad w/o EWSF                             &  V+T      & -                 & -                 & 46.7          & 64.8          & 68.5           & 51.2 \\
% VISTA(Baseline)
    \bottomrule
  \end{tabular}
  \caption{Fair comparison on E-VQA. Textual similarities are calibrated by the pool-mean ratio $R=\mu_v/\mu_t$ to align I2I/I2T scales before max-pooling across sources.}
  \label{tab:fair_comparison}
  \vskip -10pt
\end{table}

\textbf{Candidate Pool Fairness Analysis.}
Table~\ref{tab:fair_comparison} provides a controlled comparison under 
matched candidate pool sizes and analyzes the effectiveness of UniHEAR 
independent of retrieval pool scale. Within each retrieval mode, UniHEAR 
is compared with EchoSight and OMGM under identical candidate pools and 
the same untuned Qwen3-VL-8B generator. Under matched settings, UniHEAR 
achieves Recall@1 improvements of 4.2 and 1.0 points over EchoSight and 
OMGM, respectively, while requiring substantially lower entity reranking 
time. 
We further evaluate the contribution of reranking by comparing UniHEAR 
with EVA-CLIP-8B without reranking. Reranking substantially improves 
Recall@1 from 13.3 to 40.7 on V, from 19.1 to 43.8 on T, and from 21.4 
to 49.5 on V+T, demonstrating the effectiveness of the proposed 
reranking design. 
Finally, the V+T setting further analyzes the contributions of RAMG and 
EWSF. Adding RAMG improves Recall@1 from 41.6 to 46.7, while introducing 
EWSF further increases it to 49.5. The corresponding VQA accuracy also 
improves from 45.9 to 51.2 and 56.1, respectively, demonstrating that 
both components provide complementary improvements at the retrieval 
level and consistently benefit downstream VQA performance.

\subsection{Impact of Reranking Modality.}
\begin{table}[h]
    \centering
    % \small
  \begin{tabular}{lcccc}
    \toprule
    \multirow{2}{*}{\textbf{Ret. Modality}} & \multicolumn{2}{c}{\textbf{E-VQA}} & \multicolumn{2}{c}{\textbf{InfoSeek}}\\
    \cmidrule(lr){2-3} \cmidrule(lr){4-5}      & \textbf{I2I}    & \textbf{I2T}   & \textbf{I2I}   & \textbf{I2T} \\
    \midrule
    $q_t      \rightarrow c_t$                 &35.1    &25.8   &38.2   &42.0    \\
    $q_{it}   \rightarrow c_{it}$              &35.7    &37.5   &40.6   &45.5    \\
    $q_{it}   \rightarrow (c_i,c_t)$           &32.0    &27.9   &38.1   &44.3   \\
    \rowcolor{LightBlue} \textbf{$q_{it}   \rightarrow (c_{it},c_i,c_t)$}    &\textbf{39.6}    &\textbf{41.7}   &\textbf{49.9}   &\textbf{54.3}    \\
    \bottomrule
  \end{tabular}
  \caption{Modality ablation for entity-level reranking on Recall@1. Our three-modality fusion achieves optimal performance on both sources.}
  \label{tab:reranking_modality}
  \vskip -10pt
\end{table}
Table~\ref{tab:reranking_modality} ablates modality configurations
for entity-level reranking, providing direct evidence for the
modality redundancy motivating RAMG. For I2I candidates, text-only
matching performs nearly on par with full IT2IT matching on E-VQA, indicating that the already-exploited visual
modality contributes little additional discriminative signal once a
candidate has been confirmed via visual retrieval. Conversely, for
I2T candidates, text-only matching drops sharply from 37.5 
to 25.8, showing that textual features alone are largely
uninformative once a candidate has already been confirmed via
textual retrieval, leaving the complementary visual modality as the
primary discriminative source. This asymmetric pattern, most
pronounced on E-VQA's large-scale knowledge base, confirms that
reranking discriminability is systematically constrained by which
modality coarse retrieval has already exploited. Our three-modality
fusion $q_{it}{\rightarrow}(c_{it},c_i,c_t)$ achieves the best
performance across all settings, validating that each modality
contributes non-redundant evidence for heterogeneous-source
reranking.

\subsection{Impact of Hyperparameters.}
\label{sec:hyperparam_ablation}
\begin{figure}[h]
  \centering
  \includegraphics[width=\columnwidth]{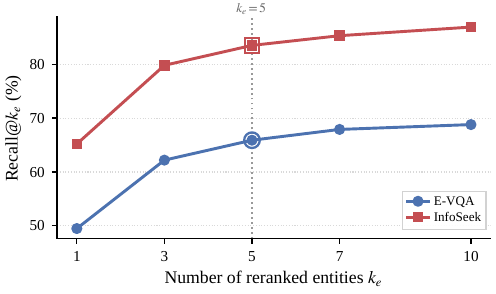}
  \caption{Effect of the number of reranked entities $k_e$ on entity
  coverage (Recall@$k_e$). Gains diminish sharply beyond the adopted
  default $k_e{=}5$ (dotted line) on both datasets.}
  \label{fig:ke_ablation}
\end{figure}

\textbf{Impact of $k_e$.}
Figure~\ref{fig:ke_ablation} reports entity coverage as a function 
of the number of reranked entities $k_e$ passed to the generation 
stage. Recall@$k_e$ increases consistently with $k_e$ on both 
benchmarks, but with rapidly diminishing returns: on E-VQA, recall 
improves by 16.4 points from $k_e{=}1$ to $k_e{=}5$, but only 
by 2.9 points from $k_e{=}5$ to $k_e{=}10$. A similar pattern 
holds on InfoSeek, where the gain from $k_e{=}5$ to $k_e{=}10$ 
is 3.4 points compared to 18.3 points from $k_e{=}1$ to $k_e{=}5$. 
We adopt $k_e{=}5$ as the default, achieving strong entity coverage 
while limiting the amount of potentially noisy context passed to 
the generator.

\begin{table}[!h]
    \centering
    \setlength{\tabcolsep}{5pt}
    \begin{tabular}{ccccccc}
        \toprule
        \multirow{2}{*}{$k_s$} & \multicolumn{6}{c}{$\alpha$} \\
        \cmidrule(lr){2-7}
        & 0.0 & 0.2 & 0.4 & 0.6 & 0.8 & 1.0 \\
        \midrule
        1 & 54.0 & \textbf{54.1} & 53.8& 53.6 & 53.5& 53.3\\
        3 & 61.4 & \textbf{61.4} & 60.9 & 60.6 & 60.3 & 59.8\\
        5 & 64.0 & \textbf{64.3} & 64.1 & 63.9 & 63.8 & 63.5\\
        7 & 65.2 & \textbf{65.3} & 65.2 & 65.2 & 65.1 & 65.0 \\
        \bottomrule
    \end{tabular}
    \caption{GT section coverage on E-VQA under varying $k_s$ and $\alpha$, 
    measured by Recall@$k_s$ per entity. $k_s$ controls the number of 
    sections selected per entity; $\alpha$ balances section-level 
    IT2T similarity against entity-level reranking score.}
    \label{tab:ks_alpha}
    \vskip -10pt
\end{table}
\textbf{Impact of $k_s$ and $\alpha$.}
Table~\ref{tab:ks_alpha} reports GT section coverage on E-VQA 
across varying $k_s$ and $\alpha$. Coverage increases consistently 
with $k_s$, confirming that allocating more sections per entity 
improves the probability of including the relevant knowledge passage. 
The gains diminish beyond $k_s{=}5$, where coverage improves by 
only 1.0 points from $k_s{=}5$ to $k_s{=}7$, suggesting that 
$k_s{=}5$ offers a favorable trade-off between coverage and 
context quality. 
Across all $k_s$ values, $\alpha{=}0.2$ consistently achieves the 
best or near-best coverage, indicating that a small weight on 
IT2T section similarity provides a useful refinement over pure 
entity-level scoring, while over-relying on section-level similarity 
alone degrades performance. We therefore adopt $\alpha{=}0.2$ 
as the default.

\begin{table}[h]
\centering
\footnotesize
\begin{tabular}{cccccccc}
\toprule
    \multirow{2}{*}{\boldmath$\tau_h$} & \multicolumn{3}{c}{\textbf{I2I}} & \multicolumn{3}{c}{\textbf{I2T}} & \multirow{2}{*}{\textbf{Discrim}$\uparrow$} \\
    \cmidrule(lr){2-4}\cmidrule(lr){5-7} & \textbf{mean} & \textbf{std} & \textbf{Degen\%} & \textbf{mean} & \textbf{std}  & \textbf{Degen\%}   & \\
    \midrule
    % 0.001                               & 0.284          &  0.450      & 60.0           & 0.261            &  0.362 & 49.6          & 0.406 \\
    % 0.002                               & 0.404          &  0.506      & 44.0           & 0.575            &  0.558 & 25.0          & 0.532 \\
    0.004                               & 0.672          &  0.618           & 23.1           & 1.224            &  0.786            &  8.7           & 0.702 \\
    0.006                               & 0.960          &  0.710           & 12.6           & 1.722            &  0.826            &  4.0           & 0.768 \\
    \rowcolor{LightBlue} 0.008          & 1.239          &  0.767           & 7.3            & 2.060            &  0.781            &  2.0           & \textbf{0.774} \\
    0.010                               & 1.491          &  0.788           & 4.5            & 2.290            &  0.707            &  1.8           & 0.748 \\
    0.012                               & 1.709          &  0.784           & 3.0            & 2.452            &  0.625            &  3.5           & 0.704 \\
    % 0.020                               & 2.279          &  0.656           & 2.0            & 2.772            &  0.342            &  26.6          & 0.499 \\
    
    % 0.050                               & 2.843          &  0.242           & 38.2           & 2.966            &  0.046            &  84.5          & 0.144  \\
    % 0.100                               & 2.960          &  0.066           & 83.2           &2.989             &  0.009 & 99.4          & 0.037 \\
\bottomrule
\end{tabular}
\caption{
Effect of entropy temperature $\tau_h$ on $H_s$. 
Degen\%: fraction of samples with a near-collapsed entropy $H_s<0.05$ or
a near-maximal entropy $H_s>2.95$, where $H_{\max}=\log20\approx3.00$.
Discrim: average standard deviation of $H_s$ across I2I and I2T, reflecting the discriminative power of $H_s$.
}
\label{tab:tau_analysis}
\vskip -20pt
\end{table}

\textbf{Impact of $\tau_h$.}
Table~\ref{tab:tau_analysis} sweeps the entropy temperature $\tau_h$ and
reports its effect on the pool-level entropy $H_s$ over 200k samples. At
$\tau_h{=}0.004$, the softmax distribution over similarity scores
becomes overly peaked, driving 23.1\% of I2I samples into the degenerate
low-entropy regime where $H_s{<}0.05$, which collapses $H_s$'s ability
to distinguish confident from ambiguous retrievals. As $\tau_h$
increases beyond 0.010, the distribution flattens toward the maximum
entropy $H_{\max}{=}\log 20{\approx}3.00$. Since I2T similarity scores
span a narrower range than I2I, I2T begins saturating into the opposite
degenerate regime first: its fraction of samples with $H_s{>}2.95$ rises
from 1.8\% at $\tau_h{=}0.010$ to 3.5\% at $\tau_h{=}0.012$.
$\tau_h{=}0.008$ achieves the best trade-off, attaining the highest
Discrim value of 0.774 while keeping degeneration low on both sources at
7.3\% for I2I and 2.0\% for I2T, and is adopted as the default.

\begin{figure}[h]
  \centering
  \includegraphics[width=\columnwidth]{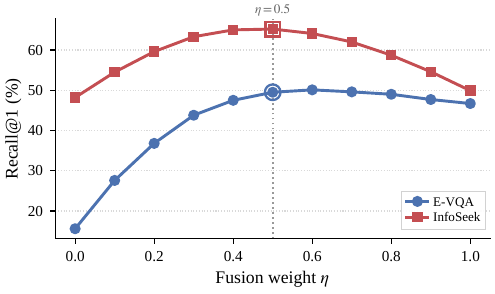}
  \caption{Effect of the entity similarity fusion weight $\eta$ on
  Recall@1. The dotted line marks the adopted default $\eta{=}0.5$.}
  \label{fig:eta_ablation}
\end{figure}

\textbf{Impact of the Score Fusion Weight $\eta$.}
Figure~\ref{fig:eta_ablation} examines the impact of the fusion weight
$\eta$. We extend the final entity ranking score (Eq.~\ref{eq:Score_Fusion}) into a
weighted score fusion $\mathrm{sim}(e){=}\eta\,\mathrm{sim}_r(e)+
(1{-}\eta)\,\mathrm{sim}_c(e)$ and sweep $\eta\in[0,1]$ on Recall@1.
Setting $\eta{=}0$ reduces ranking to the coarse retrieval score
$\mathrm{sim}_c(e)$ alone without any learned reranking signal, yielding
the lowest performance on both datasets at 15.6 on E-VQA and 48.2 on
InfoSeek. Performance rises sharply as $\eta$ increases, peaking at
$\eta{=}0.6$ on E-VQA with 50.1 and at $\eta{=}0.5$ on InfoSeek with
65.2, before gradually declining toward $\eta{=}1$, where ranking relies
solely on the learned score $\mathrm{sim}_r(e)$ without EWSF. E-VQA is
substantially more sensitive to $\eta$ than InfoSeek, spanning 34.5
points between its minimum and maximum compared to InfoSeek's
17.0-point spread. This reflects E-VQA's larger and noisier 2M-entity
knowledge base, where the coarse retrieval prior alone is markedly less
reliable. We adopt $\eta{=}0.5$ as the default, close to both datasets'
individual optima, for a better trade-off between performance and
formulation simplicity that avoids introducing an additional
hyperparameter.

\section{Case Study}
\begin{figure*}[t]
  \centering
  \includegraphics[width=0.915\textwidth]{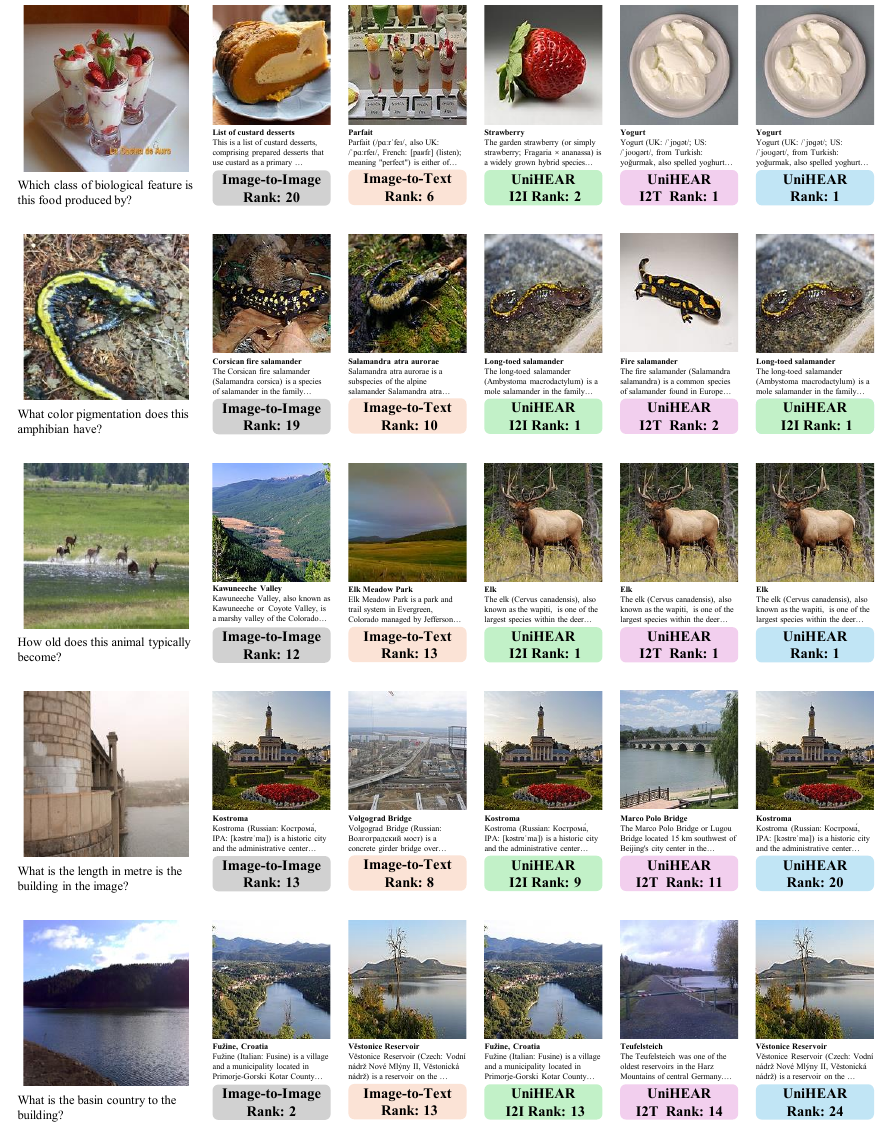}
  \caption{Qualitative retrieval examples on E-VQA and InfoSeek. Each sample shows the query and retrieval results with ground-truth ranks. Rows 1-3: Success cases where UniHEAR achieves rank 1 despite poor single-source performance. Rows 4-5: Failure cases where heterogeneous-source retrieval does not improve over coarse retrieval.}
  \label{fig:case_study}
\end{figure*}

Figure~\ref{fig:case_study} visualizes representative retrieval examples 
to provide qualitative insights into UniHEAR's behavior.

\textbf{Success Cases (Rows 1--3).}
Row 1 demonstrates that despite the ground-truth entity ranking 20th in I2I and 6th in I2T coarse retrieval, RAMG 
successfully exploits source-complementary signals to promote it to 
rank 1, confirming the value of heterogeneous-source fusion on 
knowledge-intensive queries.

Rows 2--3 show two success cases with distinct challenges. 
In Row 2, both coarse retrievers return visually similar salamander 
species, yet RAMG correctly identifies the 
ground-truth Long-toed salamander
 by leveraging complementary modality cues. 
Row 3 presents a particularly challenging scenario: the query image 
depicts an open meadow scene, causing both I2I and I2T to retrieve 
landscape-related entities rather 
than the animal in question. Crucially, the question asks about the 
animal's lifespan, and UniHEAR successfully identifies the correct 
entity by jointly reasoning over the question semantics and 
visual content, demonstrating the advantage of question-aware 
multimodal reranking over appearance-driven coarse retrieval.

\textbf{Failure Cases (Rows 4--5).}
Row 4 illustrates a failure where the final rank degrades to 20 despite 
reasonable coarse retrieval. The I2T source retrieves the correct entity at rank 8, but the high-confidence I2I signal dominates the 
Entropy-Weighted Source Fusion, suppressing the correct candidate. This 
reveals a limitation of our training-free fusion strategy: when one 
source exhibits extremely high similarity concentration, 
it can over-weight an incorrect retrieval and override complementary 
evidence from the other source.

Row 5 shows a similar pattern where EWSF assigns disproportionate 
weight to the I2T source, pushing the ground-truth entity to rank 24 despite its presence in the coarse 
candidate pool. These failure cases highlight that the unsupervised 
nature of EWSF, while effective in most settings, remains sensitive 
to extreme source confidence imbalances, suggesting that learning-based 
source reliability estimation is a promising direction for future work.

\end{document}